# Shear Stress Distribution Prediction in Symmetric Compound Channels Using Data Mining and Machine Learning Models

Zohreh Sheikh Khozani [1], Khabat Khosravi [2], Mohammadamin Torabi [3], Amir Mosavi [4,5], Bahram Rezaei [6], Timon Rabczuk [1]

[1] Institute of Structural Mechanics, Bauhaus Universität-Weimar, D-99423 Weimar, Germany

[2] Department of watershed management engineering, Faculty of natural resources, Sari agricultural science and natural resources university, Sari, Iran.

[3] Department of Civil and Environmental Engineering, Idaho State University

[4] School of the Built Environment, Oxford Brookes University, Oxford OX30BP, UK.

[5] Kando Kalman Faculty of Electrical Engineering, Obuda University, Budapest 1034, Hungary.

[6] Department of Civil Engineering, Bu-Ali Sina University, Hamedan, Iran.

**Abstract**

Shear stress distribution prediction in open channels is of utmost importance in hydraulic structural engineering as it directly affects the design of stable channels. In this study, at first, a series of experimental tests were conducted to assess the shear stress distribution in prismatic compound channels. The shear stress values around the whole wetted perimeter were measured in the compound channel with different floodplain widths also in different flow depths in subcritical and supercritical conditions. A set of, data mining and machine learning models including Random Forest (RF), M5P, Random Committee (RC), KStar and Additive Regression Model (AR) implemented on attained data to predict the shear stress distribution in the compound channel. Results indicated among these five models, RF method indicated the most precise results with the highest $R^2$ value of 0.9. Finally, the most powerful data mining method which studied in this research (RF) compared with two well-known analytical models of Shiono and Knight Method (SKM) and Shannon method to acquire the proposed model functioning in predicting the shear stress distribution. The results showed that the RF model has the best prediction performance compared to SKM and Shannon models.

# 1. Introduction

In the design of hydraulic structures; the boundary shear stress distribution is an essential factor to understand most of the flow characteristics such as the flow resistances, sediment transport, and cavitation problems. It is suggested that, the stress distribution depends on some parameters such as the flume geometry, the hydraulic condition, the boundary roughness, particularly the streamwise velocity component and the secondary flow pattern (Chiu and Chiou, 1986; Chiu and Lin, 1983; Flintham and Carling, 1988; Ghosh and Roy, 1970; Knight et al., 1994). Since the compound cross section is the nearest section to the rivers, understanding the distribution of shear stress along the periphery of compound channels is essential. Furthermore, studying the river morphology and engineering the river bed and banks is dependent on it. In addition, analysis and design of flood control structures depends on extended knowledge on the distribution of shear stresses in the flooding route. Literature includes various investigations considering different methods and case studies (Khatua and Patra, 2007; Knight and Hamed, 1984; Naik and Khatua, 2016; Rezaei and Knight, 2010; Tominaga et al., 1989). Because of the difficulty and time-consuming of direct and indirect shear stress measurement, many analytical, semi-analytical, and numerical methods have been currently developed (Shiono and Knight, 1988; Khodashenas and Paquier, 1999; Yang and Lim, 2005; Yang et al., 2012; Bonakdari et al., 2015; Sheikh Khozani et al., 2017a; Sheikh Khozani et al., 2017b). Rezaei and Knight (2009) modified the Shiono and Knight method (SKM) to predict the shear stress distribution in the compound channel with non-prismatic floodplains. Sheikh Khozani and Bonakdari (2016) compared five different analytical models to estimate the shear stress distribution in compound channels with prismatic rectangular shapes. They investigated the performance of each model in estimating shear stress in each section of the compound channel. They deducted the method of Tsallis entropy could estimate good results with fewer calculations.

Nowadays applying soft computing and data mining methods in forecasting different hydraulic and hydrology phenomena are in progress (Genç et al., 2015; Bonakdari et al., 2018; Sheikh Khozani et al., 2018a; 2018b; Azad et al., 2018; Jahanpanah et al., 2019; Sanikhani et al., 2019; Anitescu et al., 2019; Geo et al., 2019).

Nowadays applying soft computing and data mining methods in forecasting different hydraulic and hydrology phenomena are in progress. In estimating shear stress distribution Sheikh (Khozani et al., 2017) utilized the Randomize Neural Network (RNN) model in circular channels and estimated their results with results of the Shannon entropy. These researchers proposed a matrix-based equation. Khuntia et al. (2018) carried out a model of neural networks to predict the force applied to the walls in compound channel cross-sections. Sheikh Khozani et al. (2019) applied different data mining models to estimate apparent shear stress in compound channels. They deducted that by using the Bagging-M5P model the more accurate results of apparent shear stress will be obtained.

Based on the knowledge of authors there is few studies which estimated the shear stress distribution in compound channels by using data mining models. Therefore, a set of experiments were done in different flow depths and flow conditions then the extracted data was used to forecast the shear stress distribution in the smooth compound channel. About 1812 data of shear stress applied to five different models as Additive Regression (AR), M5P, KStar, Random Forest (RF), and Random Committee (RC) models. The performance of each model in prediction of the distribution of shear stress is investigated, and the most accurate model is selected. Also, the output of the most appropriate model is compared with two analytical models as Shiono and Knight (SKM) and Shannon model.

## 2. Apparatus and Proceeding of Experiments

In this study, the experiments are conducted utilizing a flume of 18m length. All experiments were performed in the flume with a simple rectangular cross-section compound channel. The flume width and depth are 1200 mm and 400 mm, respectively. The bed has a slope of $S_0 = 2.003 \times 10^{-3}$. The main channel dimensions are 398 mm, 50 , and 400 mm for width, depth, and floodplains respectively, has been constructed with PVC material. The modulus floodplain widths for the L-shaped aluminum sections in prismatic compound channels are 100 mm, 200 mm, 300 mm and 400 mm. In this study, the distribution of shear stress in the prismatic compound channel with 100 mm floodplain width is investigated (see Figs.1 and 2).

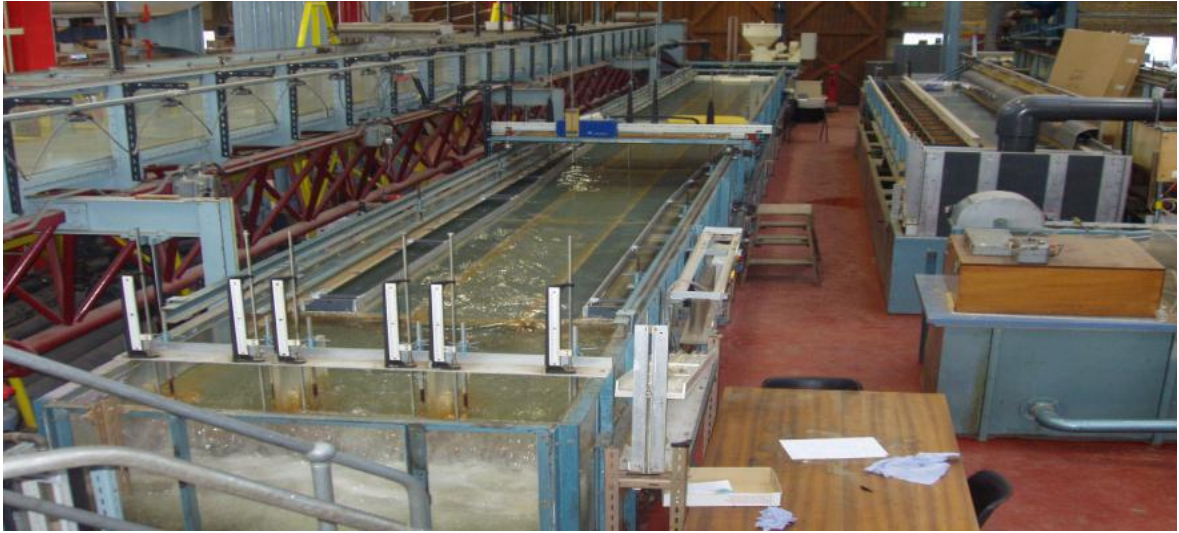

**Fig. 1.** General view of an experimental flume.

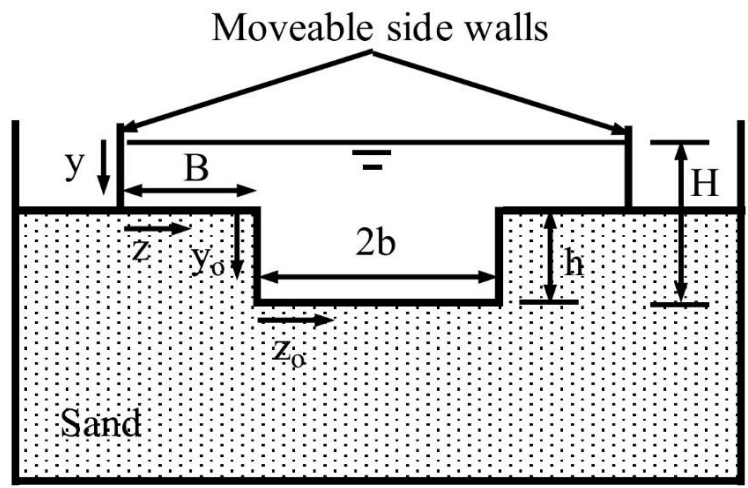

B=100-400 mm, 2b=398 mm, h=50 mm

**Fig. 2** The cross-section of prismatic compound channels with different floodplain widths.

In the expereinents, the uniform flow is controlled by a series of adjustable tailgates located in the end of the flume. OPC denotes, Overbank flow in the channel, the first three numbers after OPC refer to the floodplain width and two code numbers denoted the flow discharge. Local boundary shear stress was measured by using a Preston tube of 4.77 mm outer diameter, at the wetted channel perimeter at 25 mm transverse intervals on the bed and 10 mm vertical intervals on the walls. Note that, the above measurements were performed at one section (14 m from the channel inlet). The range of hydraulic parameters of the experimental data is presented in Table 1. The shear stress distribution was measured in different width of the floodplain.

**Table 1** The range of the main hydraulic parameters in the prismatic compound channel.

| Case | Expt. No. | H (mm) | Q (l/s) | Re×10$^{-3}$ |
|------|-----------|--------|---------|--------------|
| 1 | OPC100 | 52.78-101.50 | 12.04-39.92 | 70.77-199.45 |
| 2 | OPC200 | 52.75-104.52 | 12.03-50.03 | 49.26-175.29 |
| 3 | OPC300 | 53.26-97.37 | 12.02-50.07 | 43.21-158.58 |
| 4 | OPC400 | 53.89-93.99 | 12.02-50.10 | 34.04-128.08 |

According to the results of different research the shear stress distribution in an smooth compound channel is related to geometry of channel (the width of floodplain, $B_{fp}$, $B_{mc}$, whole channel wetted perimeter ($L$)), the transverse coordinate ($y$), bankfull depth ($h$), depth of flow over main channel ($H$), slope of channel bed ($S_0$), flow velocity ($V$), fluid density ($\rho$), gravitational acceleration ($g$) and hydraulic radius ($R$) then the dimensionless shear stress can be expressed as a function:

$$\frac{\tau}{\rho g RS} = \left( \frac{y}{L}, \frac{B_{fp}}{B_{mc}}, Fr, \frac{H}{h} \right)$$

In this study, the $y/L$, $B_{fp}/B_{mc}$, $Fr$, and $H/h$ are as input variables which applied to each model and the dimensionless shear stress is the output variable.

### 3. Material and methods

## 3.1. Data mining methods

Economist Michael Lovell who used the term "data mining" for the first time in the Review of Economic Studies (1983). Data mining is a process which discovers trends and patterns Han et al. (2011). Data mining is a subset of statistics and computer science with the mission of discovering patterns in data sets with a goal to extract trends and information from a data set and to prepare the extracted information into a required structure for further application (Witten et al., 2016).

On the other hand, in addition to the analysis step, it contains data management, inference consideration, pre-processing and post-processing of data, visualization and interestingness metrics (Khuntia et al. 2018). Data mining, unlike data analyzing, employs statistical or machine learning techniques to estimate, predict and to model patterns of the target dataset (Olson, 2007). Most common applications of data mining methods are Association learning, Anomaly detection, Cluster detection, classification, and Regression.

### 3.1.1. Random forest

Random forests (RFs) are methods for regression and classification and related tasks with constructing a multitude of decision trees. RFs considered in ensemble learning method category. This method was first introduced by Ho (1995) who implemented the stochastic discrimination to classify to the proposed by Eugene Kleinberg using the random subspace method (Barandiaran, 1998). An extension of the RFs algorithm has been registered as a trademark (Breiman, 2001). In another study by Sun et al. (2018), a new RFs algorithm has been proposed for classification based on cooperative game theory, on the other hand, the evaluation of each feature power was performed using Banzhaf power index which was traversing possible coalitions of the feature. In another study, Chen et al. (2018) proposed an adaptive variable step method based on RFs. This method from one hand was able to accelerate the training process and on the other hand, can decrease the gain of calculations of information. Based on evidence and documentation, the proposed approach was suitable to be applied in the most decision tree-based models.

In this study the optimum parameter settings of RF models including of batchsize, maximum depth of tree, number of decimal places, number execution slots, number of features, number of iterations, and number of seeds are 100, 0, 2, 1, 0, 100 and 4 respectively.

*3.1.2. M5P Model*

M5P algorithm is first introduced by Quinlan (1992). This method is the upgraded version of the M5 algorithm. Model trees can effectively handle large data sets, and in case of dealing with missing data, they are robust.

Based on Fig. 1, which shows the schematic diagram of the M5 algorithm, the process first split the input data (or input space) into subspaces.

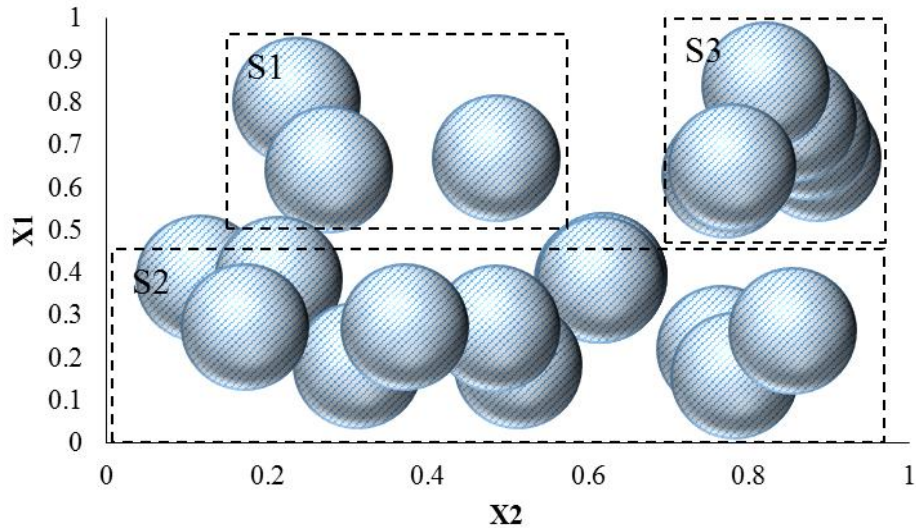

**Fig. 3.** The schematic diagram of M5 algorithm.

Figure 3 demonstrates the input space which has been divided into subspaces S1, S2, and S3. The minimization of the variation is performing by the use of linear regression approaches. After this step, in order to create a tree-like structure, information of the previous step is imported to build several nodes. In this step, the standard deviation reduction (SDR) is employed to reduce the error at the node (Eq. 1) (Wang and Witten, 1996):

$$SDR = sd(S) - \sum_i \frac{S_i}{|S|} \times sd(S_i) \qquad (1)$$

- $S$ = dataset which reaches to the node,

- $S_i$ = subspaces
- $S_d$ = the standard deviation

Lower SDR than the expected error creates over-training problems. To overcome this problem, there is a need for a smoothing process for the combination of all the models from the root to the leaf. This establishes the final model of the leaf. Finally, the resulted values of data from leaf are combined with the predicted values using linear regression for that node (Eq. 2) (Behnood et al., 2017):

$$E' = \frac{ne + ka}{n + k} \qquad (2)$$

- E'= Predicted value for the next higher node
- e = Predicted value for the current node
- a = Model prediction value
- n = Quantity of the training samples
- k = Constant value

In this paper the optimum parameter settings of M5P models including of batchsize, number of decimal places, number of instance and number of seeds are 100, 0, 2, 4, and 3 respectively.

*3.1.3. K-Star model (K\*)*

K* model or in other word K* algorithm as an Instance-based Learner and a memory-based classifier was presented by Cleary and Trigg (1995) in a conference proceedings of machine learning. The distance metric for K* technique has been performed by employing the entropy concept. Therefore, it can be claimed that the transformation probability occurs in a "random walk away" manner. Summing the probabilities classifies the K*. Generally, there is not enough evidence about how K* faces class noisy and attribute, and with the attributes mixed values in the datasets (Tejera Hernández, 2015).

In order to specify the K* technique, we have (Eq. 3 to Eq. 5):

$$0 \leq \frac{p(\bar{t}u)}{p(\bar{t})} \leq 1 \tag{3}$$

$$\sum_{u} p(\bar{t}u) = p(\bar{t}) \tag{4}$$

$$p(A) = 1 \tag{5}$$

It satisfies Eq. 6 as a consequence:

$$\sum_{\bar{t} \epsilon P} p(\bar{t}u) = 1 \tag{6}$$

Eq. 7 defines the probability function P*:

$$P^*(b|a) = \sum_{\bar{t} \epsilon P: \bar{t}(a)=b} p(\bar{t}) \tag{7}$$

The following properties have been satisfied by P*:

$$\sum_{b} P^*(b|a) = 1 \tag{8}$$

$$0 \leq P^*(b|a) \leq 1$$

Finally, the K* function will be defined as Eq. 9:

$$K^*(b|a) = -\log_2 P^*(b|a) \tag{9}$$

In this study the optimum parameter settings of KStar models including of batchsize, global blend, and minimum number of places, are 100, 1, and 1 respectively.

*3.1.4. Additive regression method*

This method is a nonparametric regression method which was first introduced by Friedman and Stuetzle (1981). This method is known as an essential part of the alternating conditional expectations algorithm. The alternating conditional expectations algorithm employs a one-dimensional smoother ($f_j(x_{ij})$ in Eq. 10) to create a class of non-parametric regression models (Eq. 10). This make the method smoother than a p-dimensional method. This technique is also more flexible compared with that for a standard linear model, but is more interpretable compared with that for a general regression surface. Multicollinearity, overfitting and model selection are consodered as application fields for an additive reggression method.

By considering $\{y_i, x_{i1}, \ldots, x_{ip}\}$, (i=1 to n) as data-set for n units, which $x_i$ indicates estimators and $y_i$ reperesents the outcome value, the additive model is as Eq. 10:

$$E[y_i|x_{i1}, \ldots, x_{ip}] = Y$$
$$= \beta_0 + \sum_{j=1}^{p} f_j(x_{ij}) + \varepsilon \qquad (10)$$

Fitting the Additive regression method can be performed by the use of the backfitting algorithm presented by. Yoshida (2018) employed a semiparametric method to explore the structure of additive regression models

The optimum parameter settings of AR models including of number of itration and shrinkage are 12 and 1 respectively.

*3.1.5. Random Committee*

Random committee belongs to the category of committee machines which works based on ensemble of predictors, e.g. ANNs, decision trees (Hwang and Hu, 2001). Thus, it is considered as an ensemble classifier which work on the basis of classification for accoplish the training. It is made using a learning mechanism which predicts the committees of the new inputs. The new imputs are generated through the integration of the estimation of every single committee members. The random committee functions as a meta-learning technique using a number of randomized classifiers. The average of estimation achieved each classifier of Random committee provides the final classification result.

Hwang and Hu (2001) documented the concept of Random Committee. He described the architecture and algorithm where some Base classifiers are constructed using a different number of random seeds. Furthermore, an estimation average generated through every base classifier form the final value for the prediction.

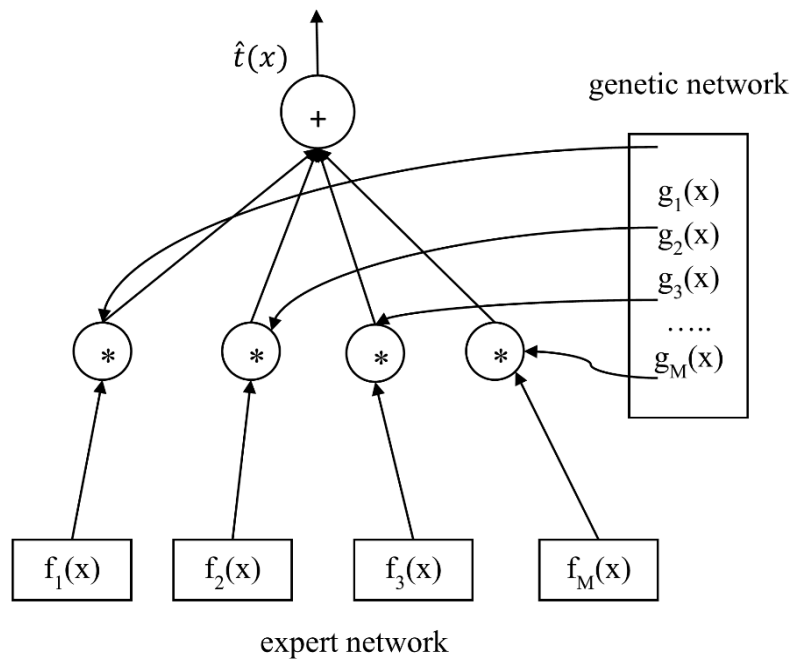

**Fig. 4.** The ME architecture. The outputs of the gating network modulate the outputs of the expert neural networks.

By assuming x as input variable and y as output variable vectors, f(x) and P(y|f(x)) will be respectively function and conditional density. By considering $X^q = \{x_1^q, \ldots, x_{NQ}^q\}$ as a set of NQ test points and let fq = (fq 1 ,...,fq NQ ) as the vector of the corresponding unknown response variables and by spliting up the input data set into M sets of data D = {$D^1$,...,$D^M$} and by denoting the data which are not in Di as $\bar{D}^i = D/D^i$ we will have in general:

$$P(f^q|\bar{D}^i, D^i) \propto P(f^q)P(\bar{D}^i|f^q)P(D^i|\bar{D}^i, f^q) \qquad (12)$$

It can be approximated Eq. 13:

$$P(D^i | \overline{D}^i, f^q) \approx P(D^i | f^q) \tag{13}$$

Now the combination of Bayes' formula and approximation generates Eq. 14:

$$P(f^q | D^{i-1}, \overline{D}^i) \approx \text{Const} \times \frac{P(f^q | D^{i-1}) P(f^q | D^i)}{P(f^q)} \tag{14}$$

approximate predictive density is calculated as Eq.15:

$$\widehat{P}(f^q | D) = \text{Const} \times \frac{\prod_{i=1}^{M} P(f^q | D^i)}{P(f^q)^{M-1}} \tag{15}$$

In this case, $\widehat{E}$ and $\widehat{cov}$ are estimated based on $\widehat{P}(f^q | D)$ as Eq. 16

$$\widehat{E}(f^q | D) = \frac{1}{C} \sum_{i=1}^{M} cov(f^q | D^i)^{-1} E(f^q | D^i) \tag{16}$$

With

$$C = \widehat{cov}(f^q | D^i)^{-1} = -(M-1)(E^{qq})^{-1} + \sum_{i=1}^{M} cov(f^q | D^i)^{-1} \tag{17}$$

The above integration of the committee members predictions ressembles the Bayesian committee machine (Hwang and Hu, 2001).

The optimum parameter settings of RC models of Batchsize, number of decimal places, number Execution slots, number of itration, and number of seed are 100, 1, 1, 15 and 1 respectively.

### *3.2. Analytical models*

### *3.2.1. SKM Model*

The Navier–Stokes equation for a fluid element in steady uniform flow can be written as:

$$\rho \left( v \frac{\partial u}{\partial y} + w \frac{\partial w}{\partial z} \right) = \rho g S_0 + \frac{\partial \tau_{yx}}{\partial y} + \frac{\partial \tau_{zx}}{\partial z} \tag{18}$$

where $S_0$ is bed slope, $u$, $v$ and $w$ are local velocities. The $\tau_{yx}$ and $\tau_{zx}$ represent the Reynolds stresses. Furthermore, $g$ and $\rho$ are gravitational acceleration and fluid density, respectively. An analytical solution for the Navier–Stokes equation to predict the lateral variation of the depth-averaged velocity in compound channels proposed earlier by Shiono and Knight (1988). It accounts for the 3D flow by the use of depth-integrated parameters to simplify its use as follow:

$$\rho g S_o H - \rho \frac{f}{8}\sqrt{1+s^2}U_d^2 + \frac{\partial}{\partial y}\left[\rho \lambda^* H^2 \sqrt{\frac{f}{8}} U_d \frac{\partial U_d}{\partial y}\right] = \frac{\partial}{\partial y}\left[H(\rho VU)_d\right] \tag{19}$$

Where $s$ is the channel side wall slope. $H$, $U_d$, $\lambda^*$, $f$, and $y$ are the local flow depth, the depth-averaged velocity, the dimensionless eddy viscosity, the Darcy–Weisbach friction factor and the lateral coordinate, respectively. Shiono and Knight (1988) proposed an analytical solution, initially ignoring the secondary flow term on the other side of the Equation (19). They concluded that by ignoring the current secondary term, the velocity profile could be determined relatively accurate. By increasing the bed friction, $f$, or the turbulent friction, $\lambda^*$, the relationship between the depth-averaged velocity and bed shear stress might be jeopardized in such a way that it became impossible to get a prediction of both profiles accurately at the same time.

Shiono and Knight (1991) proposed a secondary current model in order to improve the analytical results. From experimental results, they came to conclusion that within certain regions of the flow, the depth-averaged term on the right-hand side of differential Equation (19) varied linearly in the y-direction on the floodplains and in the main channel, in such a way, that its derivative could be replaced by the constant, $\Gamma$, in the main channel and on the floodplains. Hence

$$\Gamma = \frac{\partial}{\partial y}\left[H(\rho UV)_d\right] \tag{20}$$

$$\rho g S_o H - \rho \frac{f}{8}\sqrt{1+s^2}U_d^2 + \frac{\partial}{\partial y}\left[\rho \lambda^* H^2 \sqrt{\frac{f}{8}} U_d \frac{\partial U_d}{\partial y}\right] = \Gamma \tag{21}$$

For a flat bed region ($s \rightarrow 0$), the differential Equation (21) may be written as follow

$$\rho g H S_o - \frac{1}{8}\rho f U_d^2 + \frac{\partial}{\partial y}\left[\rho \lambda^* H^2 \left(\frac{f}{8}\right)^{1/2} U_d \frac{\partial U_d}{\partial y}\right] = \Gamma \qquad (22)$$

According to Shiono and Knight (1991), the analytical solution of Equation (22) for a prismatic compound channel with a flat bed region and vertical side walls is expressed as follows:

$$U_d = \left[A_1 e^{\gamma y} + A_2 e^{-\gamma y} + k\right]^{1/2} \qquad (23)$$

where $k = \frac{8 g S_0 H}{f}(1-\beta)$; $\gamma = \sqrt{\frac{2}{\lambda^*}}\left[\frac{f}{8}\right]^{1/4}\frac{1}{H}$ and $\beta = \frac{\Gamma}{\rho g S_0 H}$

At an interface between selected panels, different boundary conditions can be used to determine the unknown parameters A.

Having the depth-averaged velocity, the bed shear stress can be calculated as:

$$\tau_b = \frac{\rho f U_d^2}{8} \qquad (24)$$

It should be noted that the SKM is not able to model shear stress distribution on the rectangular compound channels walls.

*3.2.2. Shannon Model*

Based on the Shannon entropy concept, (Sterling and Knight, 2002) extended equations to estimate shear stress distribution in channels. They proposed equations for predicting shear stress distribution along the wetted perimeter in the circular channel without flat bed. Also they presented equations to forecast the shear stress distribution in wall and bed of trapezoidal and circular channels with sediment separately. Sheikh Khozani and Bonakdari (2016) used these models for estimating shear stress distribution to compare with other analytical models. The suggested equations by Sterling and Knight are as bellows:

$$\tau_w = \frac{1}{\lambda_w}\ln\left[1 + \left(e^{\lambda_w \tau_{\max(w)}} - 1\right)\frac{2(y - y_w)}{P_w}\right] \qquad y_w \leq y \leq \frac{P_w}{2} \qquad (25)$$

$$\tau_b = \frac{1}{\lambda_b} \ln\left[1 + \left(e^{\lambda_b \tau_{\max(b)}} - 1\right)\frac{2(y - y_w)}{P_b}\right] \qquad \frac{P_w}{2} \leq y \leq \frac{P_w}{2} + y_w \qquad (26)$$

where $\tau_w$ and $\tau_b$ are shear stress values for wall and bed of floodplain or main channel respectively, $\tau_{\max(w)}$ and $\tau_{\max(b)}$ are the maximum shear stress values for wall and bed respectively. $P_b$ and $P_w$ is the wall and bed wetted perimeter respectively, $y_w$ is an offset taken as 5 mm in the study of Sterling and Knight (2002) and $\lambda_w$, $\lambda_b$ are the Lagrange multipliers related to wall and bed of compound channel subsections respectively which calculated as:

$$\lambda_w = \left[\frac{\tau_{\max(w)} e^{\lambda_w \tau_{\max(w)}}}{e^{\lambda_w \tau_{\max(w)}} - 1} - \rho g R S_0\right]^{-1} \qquad (27)$$

$$\lambda_b = \left[\frac{\tau_{\max(b)} e^{\lambda_w \tau_{\max(b)}}}{e^{\lambda_b \tau_{\max(b)}} - 1} - \rho g R S_0\right]^{-1} \qquad (28)$$

Which $\rho$ is the fluid density, $g$ is the gravity acceleration, $R$ is the hydraulic radius and $S_0$ is the channel slope. In order to compute the maximum shear stress distribution, the proposed relations by Knight et al. (1994) these equations were utilized in studies of other researchers such as Bonakdari et al. (2015), Sheikh and Bonakdari (2015), and Sheikh Khozani and Bonakdari (2018).

## 4. Models performance evaluation

According to Dawson et al., (2007) using one statistical criterion is not suitable for evaluating a model. To investigate the performance of each model for estimating the shear stress distribution in compound channels four commonly used criteria were utilized. These applied criteria are as coefficient of determination ($R^2$), Root Mean Square Errors (*RMSE*), Mean Absolute Error (*MAE*), Nash-Sutcliffe Efficiency (*NSE*), and *BIAS*. These statistical indexes are calculated as:

$$R^2 = \frac{\left(n\sum_{i=1}^{n} x_{ip} x_{io} - \sum_{i=1}^{n} x_{ip} \sum_{i=1}^{n} x_{io}\right)^2}{\left(n\sum_{i=1}^{n} x_{ip}^2 - \left(\sum_{i=1}^{n} x_{ip}\right)^2\right)\left(n\sum_{i=1}^{n} x_{io}^2 - \left(\sum_{i=1}^{n} x_{io}\right)^2\right)} \qquad (29)$$

$$RMSE = \sqrt{\frac{\sum_{i=1}^{n}(x_{ip} - x_{io})^2}{n}} \qquad (30)$$

$$MAE = \frac{1}{n}\sum_{i=1}^{n}|x_{ip} - x_{io}| \qquad (31)$$

$$BIAS = \frac{\sum_{i=1}^{n} x_{ip} - x_{io}}{n} \qquad (32)$$

$$NSE = 1 - \frac{\sum_{i=1}^{n}(x_{ip} - x_{io})^2}{\sum_{i=1}^{n}(x_{io} - \bar{x}_{io})^2} \qquad (33)$$

which $x_{ip}$ is the predicted shear stress values by models, $x_{io}$ is the observed shear stress values in the laboratory, $\bar{x}_{io}$ and $\bar{x}_{ip}$ are the mean value of shear stress values which observed and predicted respectively and $n$ is the number of samples.

These indexes were used by Sheikh Khozani et al. (2019) to investigate the model performances in modeling apparent shear stress in compound channels.

## 5. Results and discussion

### 5.1. Selection the best statistical model

All five mentioned models were applied to shear stress distribution data which was measured in a straight rectangular compound channel. About 1812 data was used in the modeling procedure that 70% were used for the training stage and 30% for the testing stage. The results of the testing stage are shown in Figure 5 as a scatter plot and hydrograph. According to the results of this figure, the Additive Regression Model predicted the worst results of shear stress distribution with $R^2$ of 0.6745. As seen in Figure 5 the Additive Regression Model predicted the same values of shear stress in different $y/P$ in each test. Also based the results of hydrograph this model could not able estimate shear stress in the whole wetted perimeter. The M5P and KStar models show the same results to somewhat. As seen in hydrograph these models are weak in predicting the maximum and minimum shear stresses in walls and beds of main channel and floodplains, but for other $y/P$ they show more accurate results than the Additive Regression Model. The RC and RF models' predictions for the maximum and minimum shear stress values are better than those of other models. It clearly is seen from the scatter plot of Figure 5 that the RF Model with $R^2$ of 0.9003 demonstrated the most precise results than the AR, KStar, M5P, and RC models. Therefore, the predictions of the RF Model will be compared with two mentioned analytical models (the SKM and the Shannon models) in the next section.

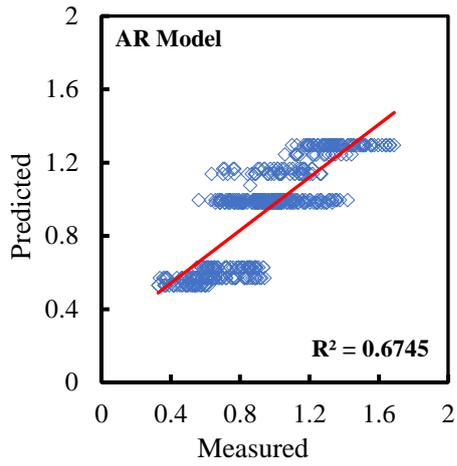 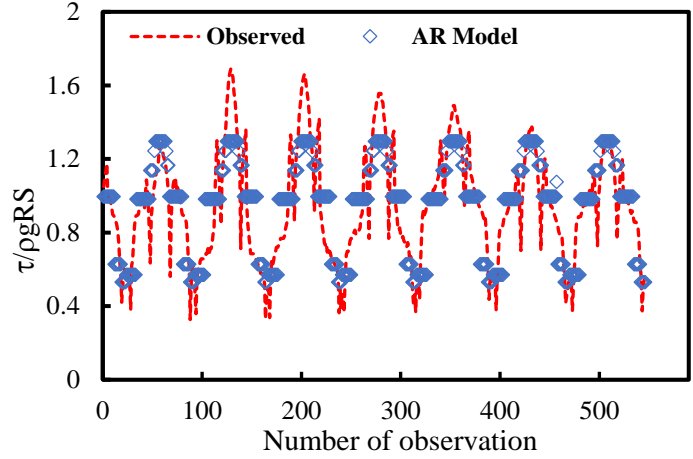

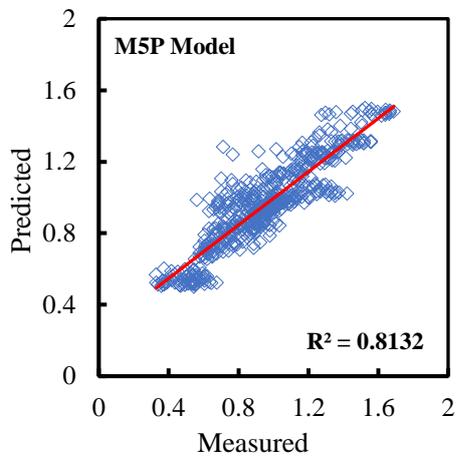 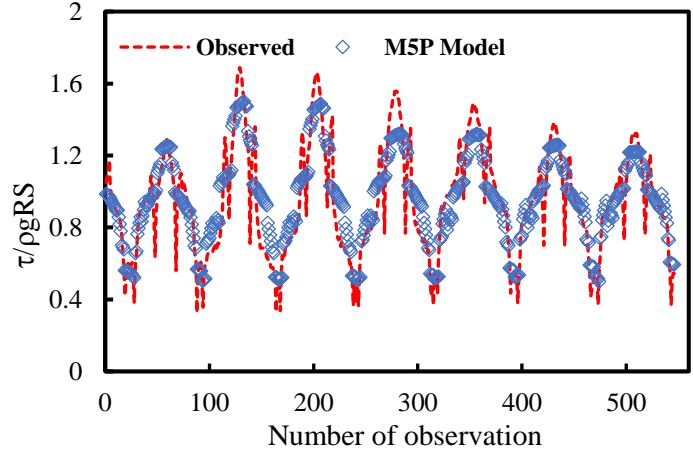

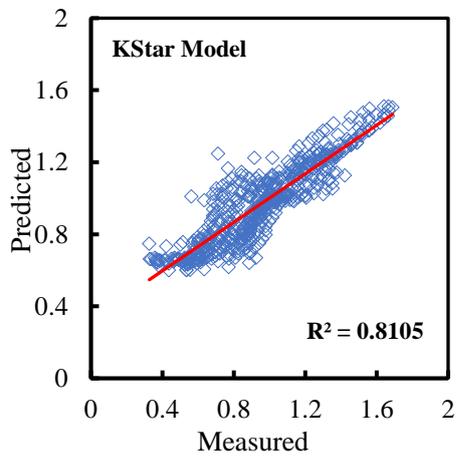 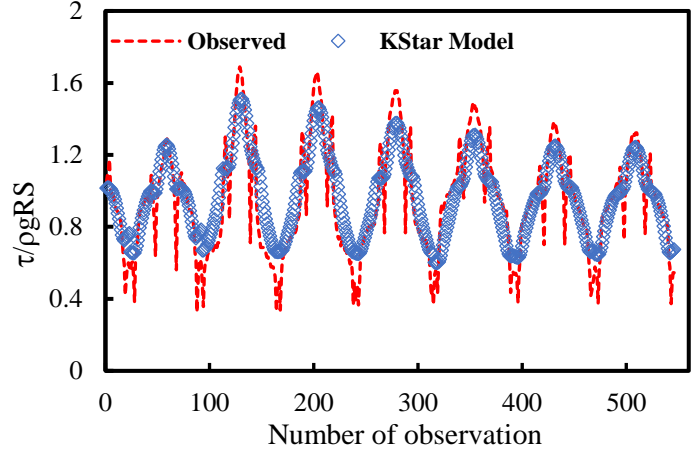
17

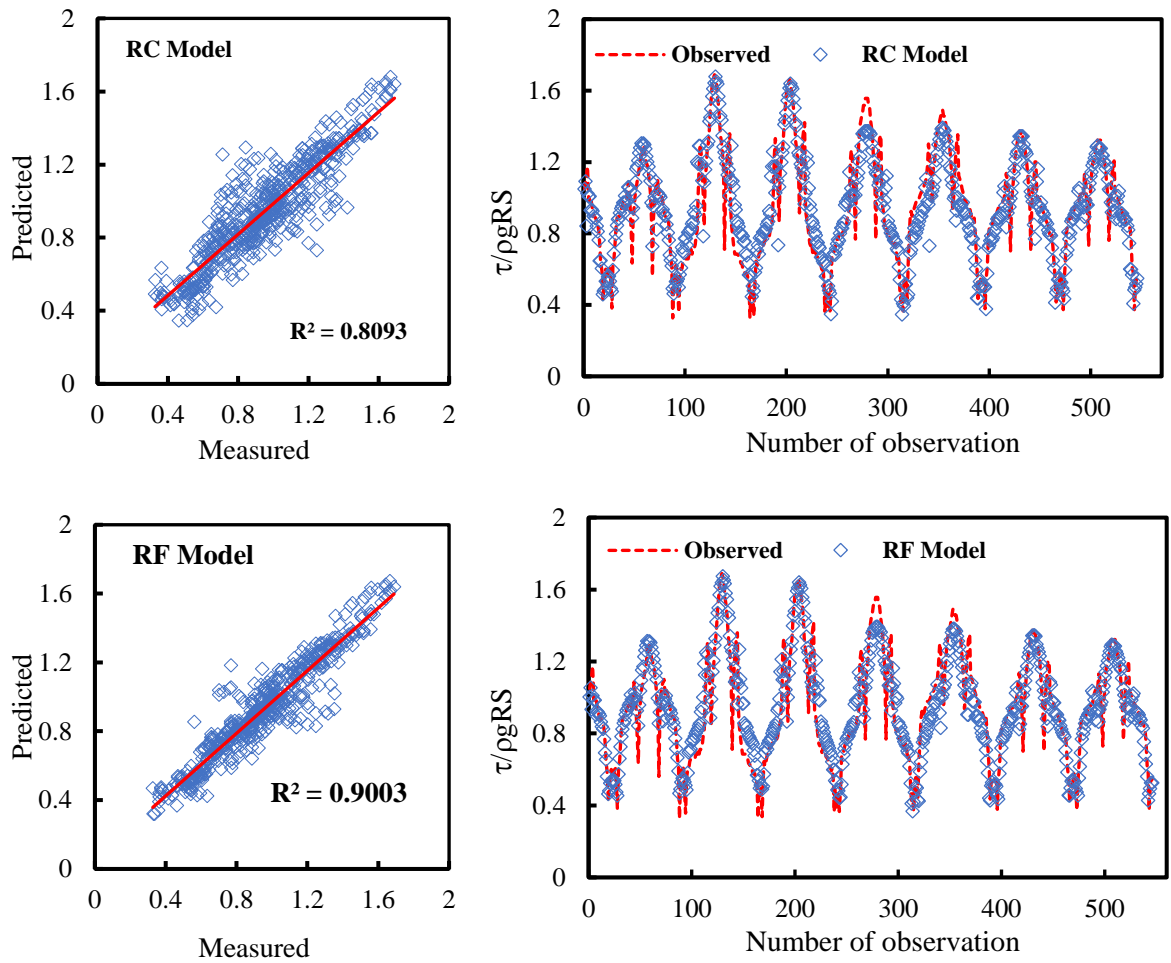

**Fig. 5.** The results of predicted shear stress values by data mining models in the testing period as scatterplots and hydrographs.

The results of statistical criteria for comparing all five data mining models are presented in Table 2. As seen in this table the performance of RF model is superior than those of other models with the lowest *RMSE* of 0.971. In addition, the AR model demonstrated the worst results for estimating shear stress distribution in compound channels with *RMSE* of 0.1707. based the results of Figure 5 and Table 2 the RF model was selected as the best model between all mentioned models to obtain the most accurate prediction values of shear stress distribution in compound channels.

**Table 2** Statistical parameters in the comparison between the soft computing methods.

| Models | RMSE | MAE | NSE | BIAS |
|--------|------|-----|-----|------|
| AR | 0.1707 | 0.1322 | 0.6697 | 0.0107 |
| M5P | 0.1305 | 0.1003 | 0.8068 | -0.0085 |
| KStar | 0.1381 | 0.1091 | 0.7838 | -0.0182 |
| RC | 0.1301 | 0.0956 | 0.8079 | 0.0055 |
| RF | 0.0971 | 0.0673 | 0.8931 | 0.0249 |

*5.2. Comparison of the models*

To estimate the shear stress distribution in a prismatic compound channel with rectangular cross-section five different data mining methods were investigated. Based on the results the RF model performed superior to those of other models in all subsections of the compound channel. In this section, the performance of the RF model is compared with the ability of the Shannon and SKM models in forecasting the shear stress distribution. Figure 6 demonstrates the comparison between two analytical models and the RF model. As seen in Figures 6a and b the SKM model shows better performance in predicting the shear stress in the bed of the main channel than the bed of floodplains. As we know the SKM model only can estimate the bed shear stress and this model is not able to predict wall shear stresses. Based on the results of Fig. 6 using the SKM model overestimated values obtain for bed shear stress of the main channel and underestimated values calculate for the shear stress of bed of floodplains. With increasing the width of floodplains, the accuracy of the SKM model predictions for the bed of the main channel was decreased.

On the other hand, in higher floodplain width the shear stress predictions values for the bed of floodplain are more precise. Also, when the width of floodplain increased, the SKM model estimates the pattern of shear stress for the bed of floodplain with higher accuracy as seen in Figures 6e, f, g, and h. The performance of the Shannon model is better than the SKM model. In all sub-sections, the Shannon model predictions are overestimated, but this model performs better for estimating wall shear stress than bed shear stress. When the width of floodplains is equal to 100 mm, the performance of the Shannon model is the same as the SKM model for main channel bed shear stress to somewhat. With increasing the width of floodplains, the results of the SKM model become weaker than the Shannon model. Between three mentioned models the RF model illustrates the best results with higher accuracy as seen in Fig 6. By using the RF model in addition to the most accurate predictions of shear stress distribution in the whole wetted perimeter, the model could estimate the pattern of shear stress distribution very well. In modeling with the RF model only using the hydraulic parameters of channel as $y/L$, $Fr$, $H/h$ and $B_{fp}/B_{mc}$ the shear stress values can estimated in whole channel boundary while in the Shannon entropy it needs to compute the Lagrange multiplier and the results are not accurate as the RF model. In addition in the SKM model we can only estimate the bed shear stress and it needs to calculate the average depth velocity and computing the shear stress needs to time-consuming procedure.

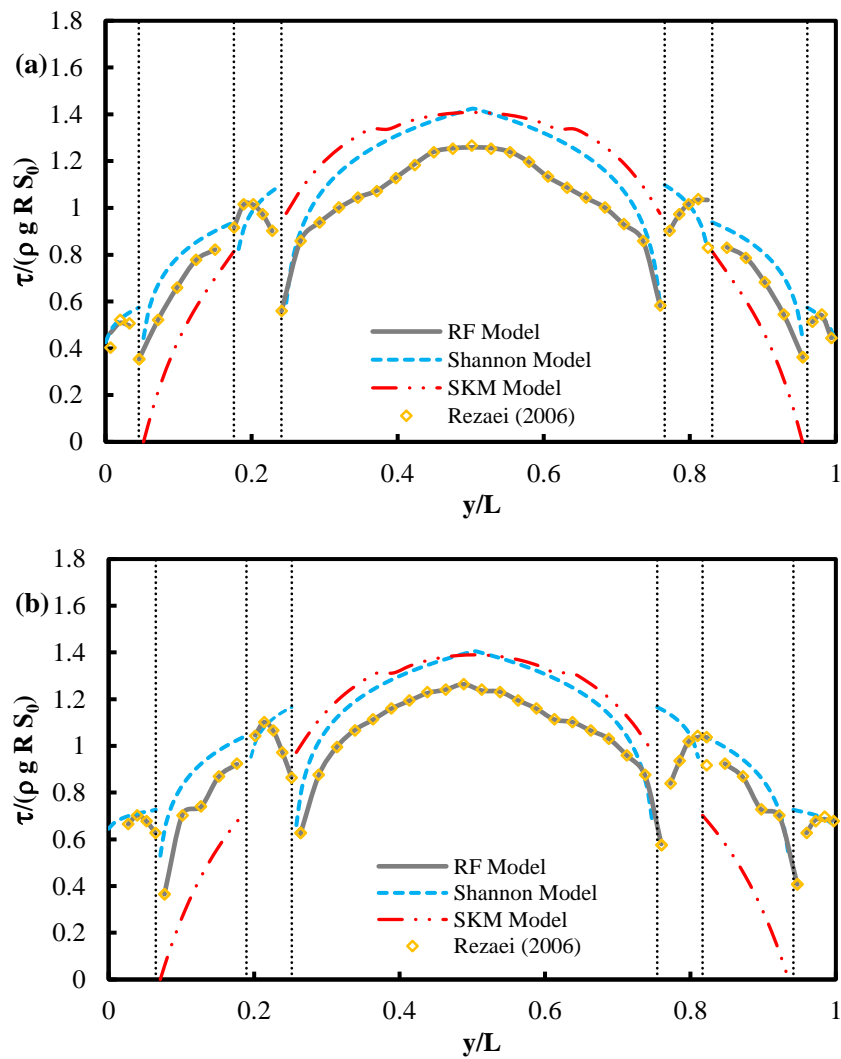

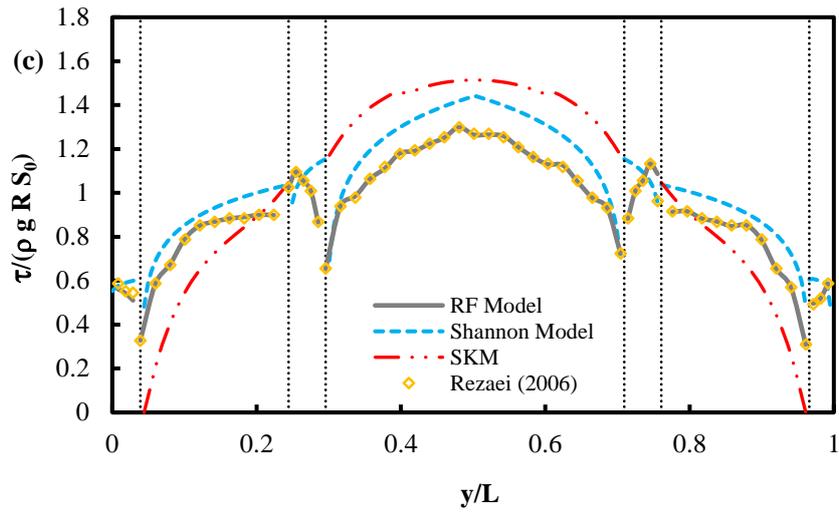

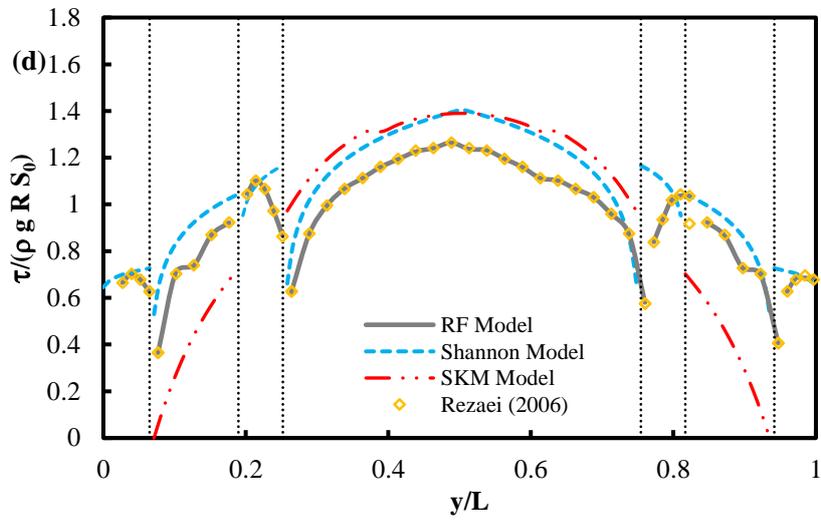

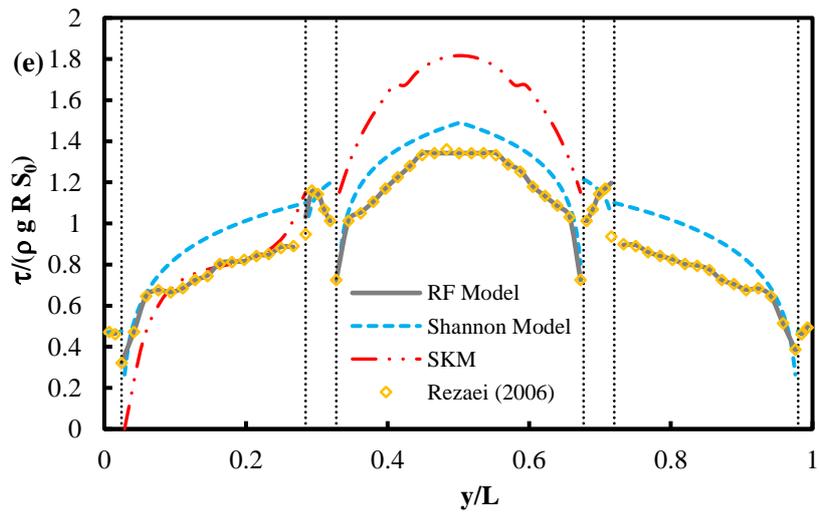

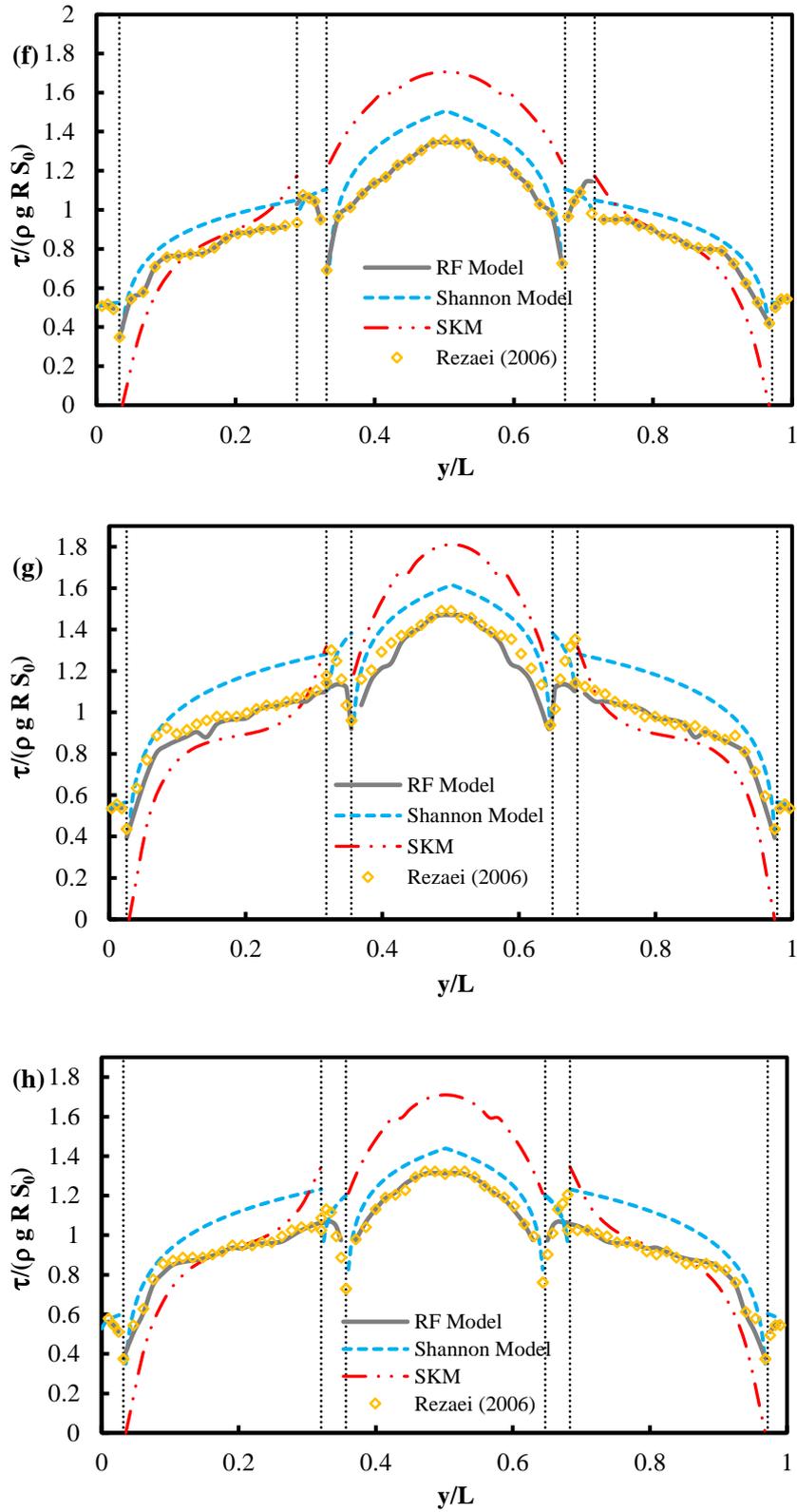

**Fig. 6.** The shear stress distribution prediction in the compound channel by RF, Shannon and SKM models for (a) OPC 100-30, (b) OPC 100-40, (c) OPC 200-35, (d) OPC 200-45, (e) OPC 300-30, (f) OPC 300-40, (g) OPC 400-40, and (h) OPC 400-50. (Rezaei 2006).

The statistical results of comparison between the RF, Shannon and SKM models are tabulated in Table 3. As we know, the lower values of *RMSE* and *MAE* indexes shows the higher performance of models to forecast a specific phenomenon. As mentioned before the SKM model predict bed shear stress of floodplains and the main channel, in Table 3 the results of the SKM model contains only these predictions. According to the results of this table, the RF model with lower values of *RMSE* and *MAE* indicates the best results of estimating shear stress distribution in compound channels. The Shannon entropy model performs better than the SKM model in predicting shear stress values. The values of *NSE* demonstrates the performance of model which graded as very good for 0.75 <*NSE*≤ 1, good for 0.65 <*NSE*≤ 0.75, satisfactory for 0.5 <*NSE*≤ 0.65, acceptable for 0.4 <*NSE*≤ 0.5, and unsatisfactory for *NSE* ≤0.4. As seen in Table 3 for the RF model the obtained values of *NSE* are higher than 0.95, therefore, the RF model has a perfect grade for estimating shear stress values. For estimating shear stress distribution values in OPC-100, OPC-200, OPC-300, and OPC-400 the results of the RF model are most precise with *RMSE* of 0.0166, 0.0255, 0.0338, and 0.0518 respectively in comparison with the Shannon and the SKM models. All in all, based the results of Figure 6 and Table 3 the RF model is the most robust model between mentioned models in this study for estimating shear stress distribution in compound channels. It is worth addition that $R^2$, *RMSE*, *MAE*, *NSE*, and *BIAS* which are used to estimate how good regression models are, in some cases, they can overestimate (or underestimate) the training data. To overcome these issues (overestimation and underestimation), Bayesian methods can be used to improve the regression model (Vu-Bac et al. 2014, 2015, 2016).

**Table 3** Statistical parameters in the comparison between the RF, Shannon and SKM models.

| **Models** | **Cases** | *RMSE* | *MAE* | *NSE* | *BIAS* |
|---|---|---|---|---|---|
| RF | OPC-100 | 0.0166 | 0.0040 | 0.9935 | 0.0022 |
|  | OPC-200 | 0.0255 | 0.0078 | 0.9877 | 0.0061 |
|  | OPC-300 | 0.0338 | 0.0084 | 0.9838 | 0.0061 |
|  | OPC-400 | 0.0518 | 0.0305 | 0.9553 | 0.0056 |
| Shannon | OPC-100 | 0.2069 | 0.1638 | 0.4966 | 0.1374 |
| | OPC-200 | 0.0938 | 0.0737 | 0.8703 | 0.0604 |
| | OPC-300 | 0.1244 | 0.1047 | 0.8291 | 0.0808 |
| | OPC-400 | 0.1350 | 0.1065 | 0.7462 | 0.1053 |
| SKM (Just for BFP and BMC) | OPC-100 | 0.2274 | 0.2008 | 0.6619 | 0.0935 |
| | OPC-200 | 0.2462 | 0.2165 | 0.6425 | 0.0947 |
| | OPC-300 | 0.2969 | 0.2207 | 0.5790 | 0.1779 |
| | OPC-400 | 0.2425 | 0.1870 | 0.6250 | 0.1301 |

## 6. Conclusion

In this research, the authors have investigated on shear stress distribution on the compound channel. Series of experiments were performed in prismatic simple rectangular cross-section compound channels of floodplain 100 mm, 200mm, 300mm and 400 mm width using flume of the University of Birmingham. The results have used for five different data mining method to predict the shear stress distribution; AR, M5P, KStar, RC and RF models. The AR model with $R^2$ of 0.6745 was not able to estimate shear stress in whole wetted perimeter accurately. The M5P and KStar models did not show appropriate results in predicting the maximum, and minimum shear stresses in walls and beds of main channel and floodplains, however for other locations of perimeter they showed more accurate outcomes rather than the AR model. The maximum and minimum shear stress values can be predicted better with the RC and RF models in comparison with the other models. The RF Model can predict the results with $R^2$ of 0.9003 which is the most precise prediction among other statistical models. Shannon and SKM analytical model have been compared with RF model, the SKM model is able to predict bed shear stress of floodplains and the main channel better than wall shear stresses, however, Shannon model can predict wall shear stresses more accurately. The accuracy of the SKM model predictions for the main channel bed decreases by increasing the floodplains width. The shear stress predictions values for the floodplain bed are more meticulous in broader floodplains. The results showed that the RF machine learning model has the lower values of *RMSE* and *MAE* in comparison with the two famous accurate analytical models' prediction of shear stress distribution in the whole wetted perimeter. Random Forest modeling technique can estimate the shear stress values in whole channel boundaries using the hydraulic parameters of $y/L$, $Fr$, $H/h$ and $B_{fp}/B_{mc}$ while, Lagrange multiplier and average depth velocity is needed in the Shannon entropy, and SKM model, respectively and the results are not as accurate as the RF model.